\documentclass[12pt]{article}

\usepackage{graphicx}
\usepackage{hyperref}
\usepackage{mathrsfs}
\usepackage{latexsym}
\usepackage{amsfonts}
\usepackage{amsmath}
\usepackage{esint}
\usepackage{amsxtra}

\def \m{\mbox}
\def \be{\begin{equation}}
\def \ee{\end{equation}}
\def\beq{\begin{eqnarray}}
\def\eeq{\end{eqnarray}}
\def \ba{\begin{array}}
\def \ea{\end{array}}
\def\sin{\mbox{sin}}
\def\cos{\mbox{cos}}

\def\cosh{\mbox{cosh}}

\def\nn{\nonumber}

\def \f{\frac}

\def \p{\partial}

\begin{document}
\vspace*{-.6in}
\thispagestyle{empty}
\baselineskip = 18pt

\vspace{.5in}
\vspace{.5in}
{\LARGE
\begin{center}
Mathieu equation and Elliptic curve
\end{center}}

\vspace{1.0cm}

\begin{center}

Wei He,\footnote{weihe@nankai.edu.cn} \qquad Yan-Gang Miao\footnote{miaoyg@nankai.edu.cn} \\
\vspace{1.0cm}\emph{Department of Physics, Nankai University,
Tianjin 300071, China}
\end{center}
\vspace{1.0cm}

\begin{center}
\textbf{Abstract}
\end{center}
\begin{quotation}
\noindent We present a relation between the Mathieu equation and a
particular elliptic curve. We find that the Floquet exponent of the
Mathieu equation, for both $q<<1$ and $q>>1$, can be obtained from
the integral of a differential one form along the two homology
cycles of the elliptic curve. Certain higher order differential
operators are needed to generate the WKB
expansion. We provide a fifth order proof.\\ \\
\end{quotation}

\newpage

\pagenumbering{arabic}

\section{Introduction}

Mathieu equation was first introduced by E. Mathieu when he studied
vibrating elliptical membranes\cite{mathieu}. Its canonical form is
\be \f{d^2u}{dz^2}+(\lambda-2q\cos 2z)u=0.\label{matheq}\ee The
related modified Mathieu equation is obtained by $z\to iz$: \be
\f{d^2u}{dz^2}-(\lambda-2q\cosh 2z)u=0.\label{modmatheq}\ee The
Mathieu equation is useful in various mathematics and physics
problems. As an example, the separation of variables for the wave
equation in the elliptical coordinates leads to the Mathieu
equation.

According to the  Floquet theory, the solution of the Mathieu
equation can be written in the form: \be u_{\nu}(z)=e^{i\nu
z}f(z).\label{mathieufun}\ee where $f(z)$ is a function of period
$\pi$, and in general $\nu$ is a constant independent of $z$. $\nu$
is called the {\em Floquet characteristic exponent}, it is a
function of the constants $\lambda$ and $q$. A classical result is
that the Floquet exponent can be obtained through the Hill's
determinant. Moreover, if $\nu$ is an even integer, then the
solution $u(z)$ is a periodic function of period $\pi$; if $\nu$ is
an odd integer, then the solution $u(z)$ is a periodic function of
period $2\pi$. In our discussion in this paper, $u(z)$ is not
required to be periodic.

The Mathieu equation has been studied for a long time, for the
collections of classical results see nice references
\cite{McLachlan, Erdelyi, WangGuo, AbramowitzStegun}, and more
recent studies in \cite{Mathieurefrecent1, Mathieurefrecent2}.

Another object we study here is a particular elliptic curve.
Geometrically the elliptic curve is topologically equivalent to a
torus, it is a Riemann surface of genus $g=1$. The relation between
the Mathieu equation and the elliptic curve naturally aries in the
integrable theory. The (modified) Mathieu equation is the
Shr\"{o}dinger equation of the two body Toda system, while the
elliptic curve is just the spectral curve of the classical Toda
system. See \cite{HokerPhong} for relevant backgrounds. As an
illustration, let us start from the
Mathieu operator\beq \mathcal {L}&=&d_z^2+\lambda-2q\cos2z\nn\\
\quad&=&d_z^2+\lambda-q(e^{i2z}+\f{1}{e^{i2z}}).\eeq Substituting
$d_z=x$ and $q(e^{2iz}-e^{-2iz})=y$, where $x,y$ are complex
coordinates. Then we have\be \mathcal
{L}=(x^2+\lambda)\pm\sqrt{y^2+4q^2}.\ee The relation \be
y^2=(x^2+\lambda)^2-4q^2\label{ellicur}\ee is nothing else but the
elliptic curve we are interested in.

The curve (\ref{ellicur}) has two independent conjugate cycles
$\alpha$ and $\beta$, they are canonical basis of the homology class
of the torus. According to the general theory of Riemann surfaces,
there is a holomorphic differential one form on the torus: \be
\omega=\f{dx}{y},\ee and we can construct two periods by integrating
$\omega$ along cycles $\alpha$ and $\beta$.\be
A=\oint_\alpha\omega,\qquad B=\oint_\beta\omega.\ee Then
$\tau=\f{B}{A}, \m{Im}\tau>0$ is the complex modulus of the elliptic
curve.

However, we are interested in a meromorphic one form,\be
\tilde{\omega}=\f{x^2dx}{y}.\ee It is related to $\omega$ by
$\omega=-2\f{\p\tilde{\omega}}{\p\lambda}+\f{\p}{\p x}(\f{x}{y})dx$,
the total derivative term will not contribute to contour integrals.
The reason for us to study $\tilde{\omega}$, rather than $\omega$,
is that it is directly related to the Mathieu equation. As a first
hint, let $x^2+\lambda=2q\cos2z$, then we have \be
\tilde{\omega}=\sqrt{\lambda-2q\cos2z}dz.\ee This is actually the
leading WKB (Wentzel-Kramers-Brillouin) solution of the Mathieu
equation. In the next section, we will see that they have an even
deeper connection. In physics literature, the elliptic curve
(\ref{ellicur}) is called Seiberg-Witten curve, and $\tilde{\omega}$
is the Seiberg-Witten differential\cite{SW9407}\footnote{In some
literatures, as in \cite{SW9407}, the elliptic curve is presented in
a cubic form $y^2=\prod_{i=1}^{3}(x-e_i)$; while in some other
literatures, as here, the elliptic curve is presented in a quartic
form $y^2=\prod_{i=1}^{4}(x-\tilde{e}_i)$. They are equivalent forms
of the same curve. The cubic form is obtained from the quartic one
by proper coordinates transformations and parameter redefinitions,
the result is that one of the zeros $\tilde{e}_i$ is mapped to
infinite. The cubic curve in \cite{SW9407} can be written as
$y^2=(x^2-\f{q^2}{16})(x-\f{\lambda}{8})$, after identifying
$u=\f{\hbar^2}{8}\lambda, \Lambda^2=\f{\hbar^2}{4}q$ according to
\cite{He2} and rescaling $x\to\hbar^2x, y\to\hbar^3y$.}.

The elliptic curve (\ref{ellicur}) can be viewed as a double
covering of the branched $x$-plane. There are four branch points at
$x=(i\sqrt{\lambda+2q},i\sqrt{\lambda-2q},-i\sqrt{\lambda-2q},-i\sqrt{\lambda+2q})$,
and two branch cuts run between
($i\sqrt{\lambda+2q},i\sqrt{\lambda-2q}$) and
($-i\sqrt{\lambda-2q},-i\sqrt{\lambda+2q}$). The homology cycle
$\alpha$ of the elliptic curve corresponds to the contour encircling
singularities $(i\sqrt{\lambda+2q},i\sqrt{\lambda-2q})$, and the
homology cycle $\beta$ of the elliptic curve corresponds to the
contour encircling singularities
$(i\sqrt{\lambda-2q},-i\sqrt{\lambda-2q})$. In the next two sections
we will show that, for $q<<1$ the Floquet exponent $\nu$ is given by
integrals of differential one forms along the $\alpha$ cycle on the
torus,  for $q>>1$ the $\nu$ is given by integrals of the same
differential forms along the $\beta$ cycle.

The relation between Mathieu equation and elliptic curve we present
here is found in our study in\cite{He1, He2}, about a relation
between gauge theories and quantization of integrable
systems\cite{NS0908}. It suggests us to develop a WKB formalism to
solve the Mathieu equation, as we explain in the next section. In
this paper we try to present the problem as a differential equation
problem, for relevant physics background, see \cite{NS0908} and
\cite{He2, MM0910}, and references therein.

\section{Floquet characteristic exponents from elliptic curve}

As the first step, we rewrite the Mathieu equation in a form
convenient for WKB expansion. Suppose $q>>1$, we rewrite it as \be
\f{\epsilon^2}{2}\f{d^2u}{dz^2}+(w-\cos 2z)u=0,\label{matheq}\ee
where $\epsilon^2=\f{1}{q}, w=\f{\lambda}{2q}$. Then $\epsilon$ is a
small expansion parameter. We expand $u(z)$ as WKB series: \be
u(z)=e^{i\int_{z_0}^zp(z^{'})dz^{'}}=e^{i\int_{z_0}^z(\f{p_0(z^{'})}{\epsilon}+p_1(z^{'})+\epsilon
p_2(z^{'})+\cdots)dz^{'}}.\label{wkbseries}\ee Substituting the
series expansion (\ref{wkbseries}) into the equation (\ref{matheq})
, we can solve $p(z)$ order by order.

Of course, the requirement $q>>1$ is not always satisfied.  One may
wonder if the results we get can be applied to the case $q<<1$. As
we will see later, by suitably adjusting $\lambda$, we actually
obtain two convergent series. One series is convergent for $q>>1,
\f{q}{\nu^2}<<1$, surprisingly it is still valid for the region
$q<<1, \nu>>1$. Another series is convergent for
$q>>1,\f{\nu^2}{q}<<1$.

The first few recursive relations for $p_m$ are: \beq
p_0&=&\sqrt{2(w-\cos2z)},\qquad
p_1=\f{i}{2}(\ln p_0)^{'},\nn\\
p_2&=&-\f{1}{8p_0}[2(\ln p_0)^{''}-((\ln p_0)^{'})^2],\qquad
p_3=\f{i}{2}(\f{p_2}{p_0})^{'},\nn\\ \cdots\eeq where the prime
denotes $\f{\p}{\p z}$.

Then we extend the Mathieu equation and its periodic solution to the
complex domain associated with the elliptic curve. Then $p(z)dz$ is
a differential one form associated to the elliptic curve. Actually,
the leading order $p_0(z)dz$ is proportional to the $\tilde{\omega}$
we introduced above. We are interested in the integrals of $p(z)dz$
along the conjugate homology cycles $\alpha$ and $\beta$ on the
elliptic curve, or equivalently, along the contours encircling
$(-\f{\pi}{2}, \f{\pi}{2})$ and
$(-\f{1}{2}\cos^{-1}w,\f{1}{2}\cos^{-1}w)$ on the $z$-plane. It is
the monodromy of the Mathieu function along cycles $\alpha$ and
$\beta$ on the torus.

The leading order integrals are related to the complete elliptic
integrals of the first and second kind, the result is: \beq
\oint_\alpha
p_0(z)dz&=&\pi\sqrt{2(w+1)}F(-\f{1}{2},\f{1}{2},1;\f{2}{w+1}),\nn\\
\oint_\beta
p_0(z)dz&=&\f{i\pi}{2}(w-1)F(\f{1}{2},\f{1}{2},2;\f{1-w}{2}).\eeq As
$p_1,p_3$ are total derivatives, the contour integrals of them are
all zero \be \oint_{\alpha,\beta}p_{2m+1}(z)dz=0,\qquad m=0,1,\ee
and \beq \oint_{\alpha,\beta}
p_2dz&=&\f{1}{8\sqrt{2}}\oint_{\alpha,\beta}\f{\sin^22z-4w\cos2z+4}{(w-\cos2z)^{5/2}}dz\nn\\
&=&-\f{1}{12\sqrt{2}}\oint_{A,B}\f{\cos2z}{(w-\cos2z)^{3/2}}dz\nn\\
&=&\f{1}{12}(2wd_w^2+d_w)\oint_{\alpha,\beta}\sqrt{2(w-\cos2z)}dz,\label{2ndorder}\eeq
where $d_w=\f{d}{dw}$. We have simplified the integral by discarding
some total derivative terms, this method was first used in
\cite{MM0910}. In a similar way we find \be \oint_{\alpha,\beta}
p_4dz=\f{1}{2^5}(\f{28}{45}w^2d_w^4+\f{8}{3}wd_w^3+\f{5}{3}d_w^2)\oint_{\alpha,\beta}
p_0dz.\ee

We can proceed the same technique to obtain the differential
operators for higher order $p_m$, by discarding total derivative
terms and simplifying the expression as far as possible. We call
these differential operators {\em generating differential
operators}. Acting these differential operators on $\oint p_0dz$, we
can get higher order contour integrals, they can be written as
combinations of the hypergeometric functions by using the
formula:\be \f{d}{dz}F(a,b,c; z)=\f{ab}{c}F(a+1,b+1,c+1; z).\ee As a
demonstration, the expression for $\oint p_2dz$ can be found in
\cite{He2}, and $\oint p_4dz$ is even more lengthy. We can get
series expansions near a suitable value of $w$ from these
hypergeometric functions. However, it is much simpler to get the
series expansion of $p_0$ first and then to act the generating
differential operators on this series.

In principle, all higher order generating differential operators can
be determined by WKB relations. However it turns out that the
calculations become very involved and it is hard to determine
whether the expressions can be simplified further by discarding
total derivative terms. Based on some observation on $p_0, p_1, p_2,
p_3, p_4$, we make a conjecture for higher order differential
operators.

{\bf{\em Claim 1: In general we have \be
\oint_{\alpha,\beta}p_{2m+1}dz=0,\nn\ee \be\oint_{\alpha,\beta}
p_{2m}dz=(c_{m,m}w^md_w^{2m}+c_{m,m-1}w^{m-1}d_w^{2m-1}+\cdots+c_{m,1}wd_w^{m+1}+c_{m,0}d_w^m)\oint_{\alpha,\beta}
p_0dz,\ee where $m=0,1,2,\cdots$, and $c_{m,i}, (i=0,1,\cdots, m)$
are numerical coefficients.}}

Now we will state the relation between the monodromy of the Mathieu
function along $\alpha,\beta$ and its Floquet exponent. The
asymptotic expansions of hypergeometric function $F(a,b,c;z)$ are
quite different for $z=0,1,\infty$. For example, let us look at the
asymptotic behavior of the leading order results
$\oint_{\alpha,\beta}p_0dz$. At $w=\infty$, we have \beq
\sqrt{2(w+1)}F(-\f{1}{2},\f{1}{2},1;\f{2}{w+1})&=&\sqrt{2w}[1-\f{1}{4}(\f{1}{2w})^{2}-\f{15}{64}(\f{1}{2w})^{4}-\f{105}{256}(\f{1}{2w})^{6}+\cdots],\nn\\
\f{1}{2}(w-1)F(\f{1}{2},\f{1}{2},2;\f{1-w}{2})&=&\f{1}{\pi}\sqrt{2w}[(\m{ln}2w-2+2\m{ln}2)+\f{1}{4}(1-2\m{ln}2-\m{ln}2w)(\f{1}{2w})^{2}\nn\\
&\quad&+\f{1}{128}(47-60\m{ln}2-30\m{ln}2u)(\f{1}{2w})^{4}+\cdots].
\label{expinf}\eeq While at $w\sim1$, with $\sigma=w-1$, we have
\beq
\sqrt{2(w+1)}F(-\f{1}{2},\f{1}{2},1;\f{2}{w+1})&=&\f{4}{\pi}+\f{(1+2\ln2-\ln\sigma)\sigma}{2\pi}+\f{(3-4\ln2+2\ln\sigma)\sigma^2}{64\pi}\nn\\
&\quad&-\f{3(2-2\ln2+\ln\sigma)\sigma^3}{512\pi}+\cdots,\nn\\
\f{1}{2}(w-1)F(\f{1}{2},\f{1}{2},2;\f{1-w}{2})&=&\f{1}{2}\sigma-\f{1}{32}\sigma^2+\f{3}{512}\sigma^3-\f{25}{16384}\sigma^4+\cdots.
\eeq

It turns out that the asymptotic expansions of
$\oint_{\alpha,\beta}pdz$ which are only powers of $w$ or $\sigma$
are related to the Floquet exponent of the Mathieu equation.

{\bf{\em Claim 2: The contour integral of $p(z)$ along the
$\alpha$-cycle gives the Floquet exponent  \be
\nu=\f{1}{\pi}\oint_\alpha p(z)dz,\ee for the case $q>>1,
\f{q}{\nu^2}<<1$(or $q<<1, \nu>>1$), the hypergeometric functions
should be expanded near $\lambda>>q>>1$, i.e. $w\sim\infty$. }}

{\bf{\em  Claim 3: The contour integral of $p(z)$ along the
$\beta$-cycle  gives the Floquet exponent  \be
\nu=\f{1}{i\pi}\oint_\beta p(z)dz,\ee for the case $q>>1,
\f{\nu^2}{q}<<1$, the hypergeometric functions should be expanded
near $\lambda\sim2q$, i.e. $w\sim1$.}}

In this way, we can get the function $\nu=\nu(w,\epsilon)$ as series
expansion of $\epsilon$ and $w$. In order to obtain the eigenvalue
$\lambda$, we need to reverse the function $\nu=\nu(w,\epsilon)$ to
get $w=w(\nu, \epsilon)$.

\section{5th order proof}

In order to prove the validity of our claims, we have to show that
the asymptotic expansions of $\nu$ given by the contour integrals
are indeed the same as results known in literatures. This has been
successfully done in \cite{He2} for the first three orders
$\epsilon^{-1}p_0+\epsilon p_2+\epsilon^3 p_4$. In this section, we
will show how to determine the generating differential operators of
$p_6$ and $p_8$, following the {\bf {\em Claim 1,2,3}}, which would
be very involved for manual calculation.

Let us start from a classical result of the asymptotic expansion for
$\lambda_\nu$:\beq
\lambda_\nu&=&\nu^2+\f{1}{2(\nu^2-1)}q^2+\f{5\nu^2+7}{32(\nu^2-1)^3(\nu^2-4)}q^4\nn\\
&+&\f{9\nu^4+58\nu^2+29}{64(\nu^2-1)^5(\nu^2-4)(\nu^2-9)}q^6+\cdots\label{eigen1}\eeq
It often states in the literature that this asymptotic expansion is
valid for $q<<1$ and $\nu\ge4$. Actually, it is also valid in the
parameter region $q>>1$ and $\f{q}{\nu^2}<<1$, this makes our WKB
method applicable. Then we reverse the series (\ref{eigen1}) to
obtain the series for $\nu$ as a function of $\lambda, q$. This can
be easily achieved with the help of computer programs, for example
the Mathematica software. We can trust the inverse results up to the
order $q^6$.

The inverse series gives\beq
\nu&=&\sqrt{\lambda}-\f{q^2}{4}\lambda^{-3/2}
-\f{q^2}{4}\lambda^{-5/2}\nn\\
&\quad&-(\f{q^2}{4}+\f{15q^4}{64})\lambda^{-7/2}
       -(\f{q^2}{4}+\f{35q^4}{32})\lambda^{-9/2}\nn\\
&\quad&-(\f{q^2}{4}+\f{273q^4}{64}+\f{105q^6}{256})\lambda^{-11/2}-(\f{q^2}{4}+\f{33q^4}{2}+\f{1155q^6}{256})\lambda^{-13/2}\nn\\
&\quad&-(\f{q^2}{4}+\f{4147q^4}{64}+\f{5005q^6}{128})\lambda^{-15/2}-(\f{q^2}{4}+\f{8229q^4}{32}+\f{42185q^6}{128})\lambda^{-17/2}\nn\\
&\quad&-(\f{q^2}{4}+\f{65637q^4}{64}+\f{722007q^6}{256})\lambda^{-19/2}
-(\f{q^2}{4}+\f{65569q^4}{16}+\f{6294301q^6}{256})\lambda^{-21/2}\nn\\
&\quad&+\mathcal {O}(\lambda^{-23/2}).\eeq We have cut off the
$\lambda$ expansion at $\mathcal {O}(\lambda^{-23/2})$, and
discarded all the $q$ expansion terms beyond the scope of the
accuracy of (\ref{eigen1}). Rewrite the inverse series in $w$ and
$\epsilon$: \beq
\nu&=&\f{1}{\epsilon}[(2w)^{1/2}-\f{1}{4}(2w)^{-3/2}-\f{15}{64}(2w)^{-7/2}-\f{105}{256}(2w)^{-11/2}]\nn\\
&\quad&+\epsilon[-\f{1}{4}(2w)^{-5/2}-\f{35}{32}(2w)^{-9/2}-\f{1155}{256}(2w)^{-13/2}]\nn\\
&\quad&+\epsilon^3[-\f{1}{4}(2w)^{-7/2}-\f{273}{64}(2w)^{-11/2}-\f{5005}{128}(2w)^{-15/2}]\nn\\
&\quad&+\epsilon^5[-\f{1}{4}(2w)^{-9/2}-\f{33}{2}(2w)^{-13/2}-\f{42185}{128}(2w)^{-17/2}]\nn\\
&\quad&+\epsilon^7[-\f{1}{4}(2w)^{-11/2}-\f{4147}{64}(2w)^{-15/2}-\f{722007}{256}(2w)^{-19/2}].\label{p6inv}\eeq
It is straightforward to expand the integral
$\oint_\alpha(\epsilon^{-1}p_0+\epsilon p_2+\epsilon^3 p_4)dz$ at
$w=\infty$, and compare with (\ref{p6inv}). They indeed
match\cite{He1, He2}.

Interestingly, the expansion (\ref{p6inv}) is precise enough to
further determine the differential operator for $\epsilon^5p_6$.
According to our first claim, we set \be \oint_{\alpha,\beta}
p_6dz=(c_{3,3}w^3d_w^6+c_{3,2}w^2d_w^5+c_{3,1}wd_w^4+c_{3,0}d_w^3)\oint_{\alpha,\beta}
p_0dz.\label{6thorder}\ee Expanding $\oint_\alpha
p_0dz=\sqrt{2(w+1)}F(-1/2,1/2,1;2/(w+1))$ at $w=\infty$ as in
(\ref{expinf}), and acting the 6th order differential operator
(\ref{6thorder}) on the series,
we get\beq \oint_\alpha p_6dz=&-&\f{3}{8}(315c_{3,3}-70c_{3,2}+20c_{3,1}-8c_{3,0})(2w)^{-5/2}\nn\\
&-&\f{105}{32}(1287c_{3,3}-198c_{3,2}+36c_{3,1}-8c_{3,0})(2w)^{-9/2}\nn\\
&-&\f{10395}{512}(3315c_{3,3} - 390c_{3,2} + 52c_{3,1}- 8c_{3,0})(2w)^{-13/2}\nn\\
&-&\f{225225}{2048}(6783c_{3,3}-646c_{3,2}+68c_{3,1}-8c_{3,0})(2w)^{-17/2}\nn\\
&-&\f{72747675}{131072}(12075c_{3,3}-966c_{3,2}+84c_{3,1}-8c_{3,0})(2w)^{-21/2}\nn\\
&+&\cdots.\label{p6expa} \eeq In order to determine the four
coefficients $c_{3,i}$, we have to match (\ref{p6expa}) with terms
of order $\epsilon^5$ in formula (\ref{p6inv}). A crucial point is
that although there are only three nonzero terms in (\ref{p6inv}),
limited by the accuracy of (\ref{eigen1}) to the $q^6$ order, the
leading term $w^{-5/2}$ is absent in (\ref{p6inv}), this fact
enables us to determine the four coefficients in (\ref{p6expa}). We
finally arrive at\be \oint_{\alpha,\beta}
p_6dz=\f{1}{2^{6}}(\f{124}{945}w^3d_w^6+\f{158}{105}w^2d_w^5+\f{153}{35}wd_w^4+\f{41}{14}d_w^3)\oint_{\alpha,\beta}
p_0dz.\ee Terms of order $w^{-21/2}$ and higher in (\ref{p6expa})
are superfluous for the determination of $c_{3,i}$. After $c_{3,i}$
are determined, these higher order terms can be subsequently
determined, too.

The $\epsilon^7p_8$ order contour integral has five unknown
coefficients\be \oint_{\alpha,\beta}
p_8dz=(c_{4,4}w^4d_w^8+c_{4,3}w^3d_w^7+c_{4,2}w^2d_w^6+c_{4,1}wd_w^5+c_{4,0}d_w^4)\oint_{\alpha,\beta}
p_0dz.\ee In order to determine the coefficients we need at least
four terms in the $\epsilon^7$ order in (\ref{p6inv}), as the
coefficient for $\epsilon^7(2w)^{-7/2}$ should vanish. Therefore, we
need the $q^8$ order contribution for $\lambda_\nu$. Fortunately, it
has been worked out in \cite{Mathieurefrecent2}: \be
\lambda_\nu^{(q^8)}=\f{1469\nu^{10}+9144\nu^8-140354\nu^6+64228\nu^4+827565\nu^2+274748}{8192(\nu^2-16)(\nu^2-9)(\nu^2-4)^3(\nu^2-1)^7}q^8.\ee
It extends the $\epsilon^7$ order terms in (\ref{p6inv}) to\be
\epsilon^7[-\f{1}{4}(2w)^{-11/2}-\f{4147}{64}(2w)^{-15/2}-\f{722007}{256}(2w)^{-19/2}-\f{1000684685}{16384}(2w)^{-23/2}].\ee
This determines $\epsilon^7\oint p_8dz$ as \be \oint_{\alpha,\beta}
p_8dz=\f{1}{2^4}(\f{127}{4725\times2^3}w^4d_w^8+\f{13}{175}w^3d_w^7+\f{517}{63\times2^4}w^2d_w^6+\f{9539}{945\times2^3}wd_w^5+\f{15229}{135\times2^7}d_w^4)\oint_{\alpha,\beta}p_0dz.\ee

We have shown that the {\bf{\em Claim 1}} is correct up to the 5th
order. Moreover, by using {\bf{\em Claim 2}} for $q<<1$, we have
determined all the coefficients in the generating differential
operators of $p_6$ and $p_8$. Then it is straightforward to expand
$\oint_\beta pdz$ to the $\epsilon^7$ order, near $w\sim1$, to
obtain the Floquet index $\nu=\nu(w,\epsilon)$ for $q>>1$.  After
reverse the series $\nu=\nu(w,\epsilon)$ to $w=w(\nu,\epsilon)$ and
rewrite it in $\nu, \lambda, q$, we get \beq
\lambda_\nu&=&2q-4\nu\sqrt{q}+\f{4\nu^2-1}{2^3}+\f{4\nu^3-3\nu}{2^6\sqrt{q}}\nn\\
&\quad&+\f{80\nu^4-136\nu^2+9}{2^{12}q}+\f{528\nu^5-1640\nu^3+405\nu}{2^{16}q^{\f{3}{2}}}\nn\\
&\quad&+\f{2016\nu^6-10080\nu^4+5886\nu^2-243}{2^{19}q^2}\nn\\
&\quad&+\f{33728\nu^7-249872\nu^5+276004\nu^3-41607\nu}{2^{24}q^{\f{5}{2}}}\nn\\
&\quad&+\f{2403072\nu^8-24881920\nu^6+45534368\nu^4-16087536\nu^2+506979}{2^{31}q^3}\nn\\
&\quad&+\f{44811520\nu^9-620967168\nu^7+1724770656\nu^5-1152647184\nu^3+130610637\nu}{2^{36}q^{\f{7}{2}}}.\nn\\
\quad\label{eigen2}\eeq We keep only terms consistent with the
accuracy limit $\epsilon^7p_8$. This is the classical result of the
Mathieu equation for $q>>1$(See formula (20.2.30) in
\cite{AbramowitzStegun}). This finishes the proof of our {\bf{\em
Claim 3}}.

\section{Concluding remarks}

We show that the Mathieu equation is closely related to an elliptic
curve, therefore there is a geometric structure for the Mathieu
equation which is not captured by asymptotic analysis. The Floquet
exponent of the Mathieu equation can be derived from integrals of
certain differential forms along the homology cycles of the curve.
These differential forms are determined by the WKB procedure.
Integrals along each homology cycle give an asymptotic expression
for the Floquet exponent expanded at a specific point, the inverse
series gives the corresponding eigenvalue. Integrals along all
homology cycles $\alpha$ and $\beta$ give the complete asymptotic
expansions (\ref{eigen1}) and (\ref{eigen2}) for the eigenvalue.

The appearance of Riemann surfaces associated with differential
equations is quite familiar in the theory of integrable models, the
Riemann surfaces are the spectral curves of the classical integrable
system while the differential equations are the Shr\"{o}dinger
equation of the same system. The (modified) Mathieu equation we
discuss here is simply the two body Toda system, the eigenvalue
formulae (\ref{eigen1}) and (\ref{eigen2}) are states with large
quantum numbers(expanded at $\lambda>>1$) and states with small
quantum numbers(expanded at $\lambda\sim2q$), respectively\cite{He1,
He2}. In fact, the Lax matrix $L(z)$ of the classical two body
periodic Toda system is\cite{bbt} \be L(z)= \left(\begin{array}{cc}
p_1 & e^{q_1-q_2}+z^{-1}e^{q_2-q_1}\\
e^{q_1-q_2}+ze^{q_2-q_1} & p_2
\end{array}\right)
\ee where $p_1,p_2$ are momentum and $q_1,q_2$ are coordinates of
the particles, $z$ is the complex spectral parameter. The spectral
curve is obtained by the determinant \be
\m{det}(\mu\mathbb{I}-L(z))=0.\ee It's of the form \be
2t(\mu)-(z+z^{-1})=0,\ee where
$2t(\mu)=\mu^2-(p_1+p_2)\mu+f(q_1-q_2,p_1p_2)$ is a polynomial of
degree two. By a change of variables, $x=\mu, y=z-t(\mu)$, the curve
becomes \be y^2=t^2(x)-1.\ee This is the same curve as
(\ref{ellicur}) modula a further coordinates change. Now let us
substitute $\mu$ by the differential $\p_z$, then formally we can
get a second order differential equation through \be
\m{det}(\mathbb{I}\p_z-L(z)).\ee It's easy to see that after some
proper coordinates changes this is the Mathieu differential
operator. So in some sense, the Mathieu equation defines a ``quantum
torus". Actually, Talalaev has constructed the ``quantum spectral
curve" for general integrable systems along this way\cite{Talalaev}.

It is possible that the relation we present here is just a
particular case of a general picture. As an example, the spectral
curve for the two body elliptic Calogero-Moser integrable system is
an elliptic curve closely related to (\ref{ellicur})(elliptic curves
always can be written in the Weierstrass form), and its quantization
leads to the $Lam\acute{e}$ equation. See a recent discussion in
\cite{MT1006}.

Another example is the direct generalization of the case we present
here, the spectral curve for the $N$-body $A_N$ periodic Toda chain
is a hyperelliptic curve of genus $g=N-1$, with $N\ge3$. The
Gutzwiller's quantization scheme of periodic Toda chain introduces a
rather involved Bethe-like quantization condition which involves
both the Floquet exponents and the integrals of
motion\cite{Gutzwiller}. Recently in \cite{KozlowskiTeschner} the
quantization condition has been rewritten in a functional form that
only involves the Floquet exponents of the associated Hill's
determinant. We wonder if the idea presented here can be generalized
to higher genus curves and provide a solution to the eigenvalue
problem of periodic Toda chain. Note that for a $N$-particle Toda
system, there are $N-1$ independent Floquet exponents $\nu_i,
i=1,2,\cdots,N-1$(the condition $\sum_{i=1}^{N}\nu_i=0$ just reduces
the center of mass motion, or is the traceless condition for the
SU(N) group), and for the associated hyperelliptic curve there are
$2(N-1)$ independent homology cycles $\alpha_i$ and $\beta_i$. There
are also a meromorphic one form and its WKB descendants on the
curve. The differential forms involve exactly $N-1$ coefficients
$I_k,$ with $k=2,3,\cdots,N$, which are the integrals of motion of
Toda chain. Among the integrals of motions $I_2$ is interpreted as
energy while $I_{k}$ for $k\ge3$ have no physical interpretation.
There is evidence that the Floquet exponents $\nu_i$ are given by
integrals of these differential forms along the homology cycles on
the hyperelliptic curve \cite{MM0911}. This is enough to determine
the functional relations between the Floquet exponents and the
integrals of motion $\nu_i=\nu_i(I_2,I_3,\cdots, I_{N})$. There are
$2(N-1)$ asymptotic expansion points at $I_k>>1$ and at the dual
points $I_k\sim I_k^{(0)}$. The critical values $I_k^{(0)}$ are
determined by the Chebyshev polynomial \cite{DouglasShenker}. If the
$N-1$ integral cycles are chosen as $(\alpha_{\lbrace
i\rbrace},\beta_{\lbrace j\rbrace})$, satisfying $\lbrace
i\rbrace\subseteq\lbrace1,2,\cdots
 N-1\rbrace, \lbrace j\rbrace\subseteq\lbrace1,2,\cdots
 N-1\rbrace$ and ${\lbrace i\rbrace}\cup{\lbrace j\rbrace}=\lbrace1,2,\cdots,N-1\rbrace, {\lbrace i\rbrace}\cap{\lbrace j\rbrace}=\emptyset$,
 then functions $\nu_{\lbrace i\rbrace}=\nu_{\lbrace i\rbrace}(I_2,I_3,\cdots, I_{N})$ have asymptotic expansions at
$I_k>>1$, while $\nu_{\lbrace j\rbrace}=\nu_{\lbrace
j\rbrace}(I_2,I_3,\cdots, I_{N})$ have asymptotic expansions at
$I_k-I_k^{(0)}<<1$. By reversing the Floquet exponents
$\nu_i=\nu_i(I_2,I_3,\cdots, I_{N})$ we obtain the eigenvalues
$I_k=I_k(\nu_1,\nu_2,\cdots,\nu_{N-1})$.

\section*{Acknowledgments}
Y-G.M was supported in part by the National Natural Science
Foundation of China under grant No.10675061.

\end{document}